\documentclass[pra,twocolumn,superscriptaddress,floatfix,amsmath,amssymb]{revtex4} 
\usepackage[english]{babel}
\usepackage{graphicx}% Include figure files
\usepackage{dcolumn}% Align table columns on decimal point
\usepackage{bm}% bold math
\usepackage{verbatim}
\usepackage{color}
\pacs{03.67.Mn, 03.65.-w, 03.65.Yz, 04.62.+v}% PACS, the Physics and Astronomy
\newcommand{\ket}[1]{\left| {#1} \right\rangle}
\newcommand{\bra}[1]{\left\langle {#1} \right|}

\newcommand{\proj}[2]{\left| {#1} \right\rangle\!\left\langle {#2} \right|}

\usepackage{color,xspace}

\def\slashchar#1{\setbox0=\hbox{$#1$} % set a box for #1
\dimen0=\wd0 % and get its size
\setbox1=\hbox{/} \dimen1=\wd1 % get size of /
\ifdim\dimen0>\dimen1 % #1 is bigger
\rlap{\hbox to \dimen0{\hfil/\hfil}} % so center / in box
#1 % and print #1
\else % / is bigger
\rlap{\hbox to \dimen1{\hfil$#1$\hfil}} % so center #1
/ % and print /
\fi}

\begin{document}

\title{Redistribution of particle and anti-particle entanglement in non-inertial frames}

\author{Eduardo Mart\'{i}n-Mart\'{i}nez}
\address{Instituto de F\'{i}sica Fundamental, CSIC, Serrano 113-B, 28006 Madrid, Spain}

\author{Ivette Fuentes\footnote{Previously known as Fuentes-Guridi and Fuentes-Schuller.}}
\address{School of Mathematical Sciences, University of Nottingham, Nottingham NG7 2RD, United Kingdom}

\begin{abstract}
We analyse the entanglement tradeoff between particle and anti-particle modes of a Dirac field from the perspective of inertial and uniformly accelerated observers. Our results show that a redistribution of entanglement between particle and anti-particle modes plays a key role in the survival of fermionic field entanglement in the infinite acceleration limit.
\end{abstract}

\maketitle

\section{Introduction}
Understanding entanglement in non-inertial frames has been central to the development of relativistic quantum information \cite{Alsingtelep,TeraUeda2,ShiYu,Alicefalls,AlsingSchul,Adeschul,KBr,LingHeZ,ManSchullBlack,PanBlackHoles,AlsingMcmhMil,DH,Steeg,Edu2,schacross,Ditta,Hu,DiracDiscord,Edu6}. The main aim of this field is to implement quantum information tasks (such as quantum teleportation) in relativistic settings.  Quantum correlations are an important resource in most quantum information applications therefore, it has been of interest to understand how entanglement can be degraded \cite{Alicefalls,AlsingSchul,Edu3,Edu4,Edu6} or created \cite{Ball,Expanfer,Edu8,EdMi} by the presence of horizons or spacetime dynamics. 

It is well-known that the entanglement between modes of bosonic and fermionic fields is degraded from the perspective of observers moving in uniform acceleration. Interestingly,  entanglement is completely degraded in the infinite acceleration limit in the bosonic case while for fermionic fields a finite amount of entanglement remains in the limit. However, the reasons for these differences were not completely clear.  In this paper we show that a redistribution of  entanglement between particle and anti-particle modes plays a key role for the preservation of fermionic field entanglement  in the infinite acceleration limit.

In our analysis we consider entangled states which involve particle and antiparticle field modes from the perspective of inertial observers.  Previous studies considered entangled  states involving exclusively particle modes from the inertial perspective.  To study particle and antiparticle entangled states we develop a generalization of the formalism introduced in \cite{Jorma} which relates general Unruh and Rindler modes. This formalism refines the single-mode approximation \cite{Alsingtelep,AlsingMcmhMil}  which has been extensively used in the literature.  
 In particular, we will consider in our analysis a fermionic maximally entangled state which has no neutral bosonic analog.  This state which is  entangled in the particle/antiparticle degree of freedom can be produced, for example, in conjugated pair creation or in the production of cooper pairs.   The analysis of such states is only possible under the mode transformations we introduce here since the single approximation  \cite{Alsingtelep,AlsingMcmhMil}  does not hold in this case.

 Considering a more general set of states from the inertial perspective  allowed us to understand that in non-inertial frames entanglement redistributes between particle and anti-particle modes.  This is a somewhat similar effect to that observed in the inertial case:  entanglement redistributes between spin and position degrees of freedom from the perspective of different inertial observes  \cite{Peres,Lamata}.  Interestingly, one can conclude that fermionic entanglement remains finite in the infinite acceleration limit due to this redistrubution of entanglement, which does not occur in the bosonic case. Our results are in agreement with previous results which show that main differences in the behaivour of entanglement in the bosonic and fermionic case are due to Fermi-Dirac and Bose-Einstein statistics, contrary to the idea that he dimension of the Hilbert  played an important role \cite{Edu5}.

This paper is organized as follows:  in section \ref{sec1} we introduce transformations between Minkowski, Unruh and Rindler modes for fermionic fields. This section extends results from\cite{Jorma} by including anti-particle modes. In section \ref{sec2}, we analyse the entanglement transfer between the particle and antiparticle sectors in different kind of maximally entangled states when one of the observers is uniformly accelerated. Finally, conclusions and discussions are presented in section \ref{conclusions}.

\section{Dirac field states form the perspective of uniformly accelerated observers}\label{sec1}

We consider a Dirac field in $1+1$dimensions. The field can be expressed from the perspective of inertial and uniformly accelerated observers. In this section we introduce the transformations which relate the mode operators and states from both perspectives.  Such transformations have been introduced in \cite{Jorma}  for particle states. Here we extend this results including transformations for anti-particle modes which will be needed in our analysis. 

Minkowski coordinates $(t,x)$ are an appropriate choice of coordinates to express the field from the perspective for inertial observers. However, in the uniformly accelerated case Rindler coordinantes  $(\eta,\chi)$ must be employed.  The coordinate transformation is given 
 by
\begin{equation}
\label{Rindlertransformation}
\eta = \text{atanh} \! \left(\frac{t}{x}\right), 
~~~\chi = \sqrt{x^2-t^2},
\end{equation}
\begin{figure}[h]
\begin{center}
\includegraphics[width=.50\textwidth]{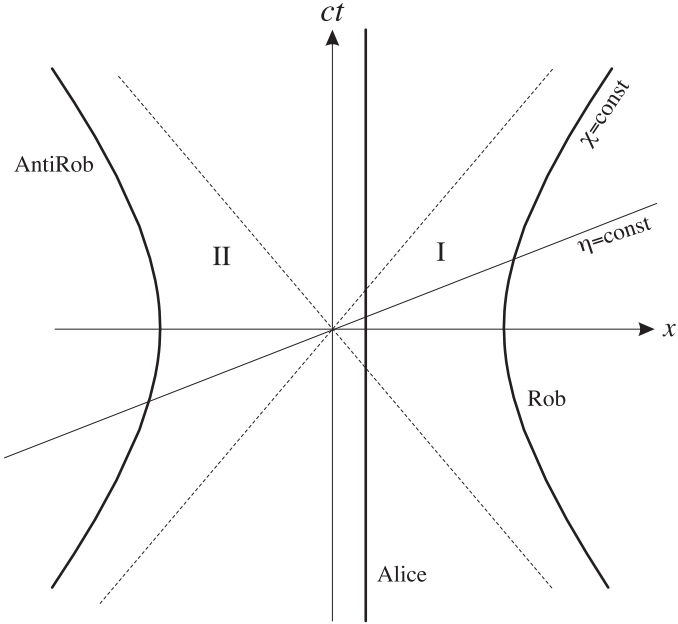}
\end{center}

\caption{ Rindler space-time diagram: lines of constant position $\chi=\text{const.}$ correspond to hyperbolae and all curves of constant  $\eta$ correspond straight lines that converge in the  origin. An uniformly accelerated observer Rob travels along a hyperbola constrained to either region I or region II.}
\label{figbosons}
\end{figure}
where $0<\chi<\infty$ and $-\infty < \eta < \infty$. The transformation is defined in two  spacetime regions $|t|<x$ 
and $x<-|t|$ called region I and II, respectively .
The curve $\chi = 1/a$, where $a$ is a positive constant
of dimension inverse length, is  the world line of a
uniformly-accelerated observer whose proper acceleration equals~$a$. The proper time of this observer is given by $\eta/a$ in region I and by
$-\eta/a$ in region~II\null.  Note that $\partial_\eta$ is a timelike Killing
vector in both I and~II, and it is future-pointing in I and 
past-pointing in~II\null.

The Dirac field $\phi$ satisfies the
equation  $\{i\gamma^{\mu}(\partial_{\mu}-\Gamma_{\mu})+m\}\phi=0$
where $\gamma^{\mu}$ are the Dirac-Pauli matrices and $\Gamma_{\mu}$ are spinorial
affine connections. The field expansion in terms of the Minkowski solutions of the Dirac equation is
\begin{equation}
\phi=N_\text{M}\sum_k\left(c_{k,\text{M}}\, u^+_{k,\text{M}}+ d_{k,\text{M}}^\dagger \ u^{-}_{k,\text{M}}\right),
\end{equation}
where $N_\text{M}$ is a normalisation constant.  The label $\pm$  denotes  positive and negative energy solutions (particles/antiparticles) with respect to the Minkowskian Killing vector field $\partial_{t}$. The label $k$ is a multilabel including energy and spin $k=\{E_\omega,s\}$ where $s$ is the component of the spin on the quantisation direction. $c_k$ and $d_k$ are the particle/antiparticle operators that satisfy the usual anticommutation rules
\begin{equation}
\{c_{k,\text{M}},c_{k',\text{M}}^\dagger\}=\{ d_{k,\text{M}},d_{k',\text{M}}^\dagger\}=\delta_{kk'},
\end{equation}
and all other anticommutators vanishing. The Dirac field operator in terms of Rindler modes (solutions of the Dirac equation in Rindler coordinates) is given by
\begin{equation}
\phi\!=N_\text{\text{R}}\sum_j\left(c_{j,\text{I}} u^+_{j,\text{I}}+ d_{j,\text{I}}^\dagger u^{-}_{j,\text{I}}+c_{j,\text{II}} u^+_{j,\text{II}}+ d_{j,\text{II}}^\dagger  u^{-}_{j,\text{II}}\!\right),
\end{equation}
where $N_\text{R}$ is again a normalisation constant.  $c_{j,\Sigma},d_{j,\Sigma}$ with $\Sigma=\text{I},\text{II}$ represent Rindler particle/antiparticle operators. Note that operators in different regions $\Sigma=\text{I},\text{II}$ do not commute but anticommute. $j=\{E_\Omega,s'\}$ is again a multi-label including all the degrees of freedom. Here  $u^\pm_{k,\text{I}}$ and $u^\pm_{k,\text{II}}$  are the positive/negative frequency solutions of the Dirac equation in Rindler coordinates with respect to the Rindler timelike Killing vector field in region $\text{I}$ and $\text{II}$, respectively. The modes  $u^\pm_{k,\text{I}}$ ($u^\pm_{k,\text{II}}$)  do not have support outside the right (left) Rindler wedge. 
The annihilation operators $c_{k,\text{M}},d_{k,\text{M}} $ define the Minkowski vacuum $\ket{0}_\text{M}$ which must satisfy
\begin{equation}
c_{k,\text{M}}\ket{0}_\text{M}= d_{k,\text{M}}\ket{0}_\text{M}=0, \qquad \forall k.
\end{equation}
In the same fashion $c_{j,\Sigma},d_{j,\Sigma}$, define the Rindler vacua in regions $\Sigma=\text{I},\text{II}$
\begin{equation}
c_{j,{{\Sigma}}}\ket{0}_\Sigma=d_{j,\Sigma}\ket{0}_\Sigma=0, \qquad \forall j, \, \Sigma=\text{I},\text{II}.
\end{equation}	
The transformation  between the Minkowski and Rindler modes is given by
\begin{align}
\nonumber u^+_{j,\text{M}}=&\sum_k\left[\alpha^\text{I}_{jk}u^+_{k,\text{I}} + \beta^{\text{I}*}_{jk} u^-_{k,\text{I}}+\alpha^\text{II}_{jk}u^+_{k,\text{II}}+\beta^{\text{II}*}_{jk} u^-_{k,\text{II}}\right].
\end{align}
\begin{align}
\nonumber u^{-}_{j,\text{M}}=&\sum_k\left[\gamma^\text{I}_{jk} u^+_{k,\text{I}} + \eta^{\text{I}*}_{jk} u^{-}_{k,\text{I}}+\gamma^\text{II}_{jk}u^+_{k,\text{II}}+\eta^{\text{II}*}_{jk} u^{-}_{k,\text{II}}\right].
\end{align}
The coefficients which relate both set of modes are given by the inner product
\begin{equation}
(u_k,u_j)=\int d^3x\, u_k^\dagger u_j,
\end{equation}
so that he  Bogoliubov coefficients yield \cite{Jauregui,Langlois,Jorma},
\begin{equation}\label{bogosolved}
\begin{array}{c}
\alpha^\text{I}_{jk}=e^{i\theta E_\Omega}\dfrac{1+i}{2\sqrt{\pi E_\omega}}\,\cos r_{\Omega}\,\delta_{s s'},\\[4mm]
\beta^\text{I}_{jk}=-e^{i\theta E_\Omega}\dfrac{1+i}{2\sqrt{\pi E_\omega}}\,\sin r_{\Omega}\,\delta_{s s'},\\[4mm]
\gamma^\text{I}_{jk}=-\beta^\text{I*}_{jk}, \qquad \eta^\text{I}_{jk}=\alpha^\text{I*}_{jk},\\[4mm]
\alpha^\text{II}=(\alpha^\text{I})^*\!\!\qquad \beta^\text{II}=(\beta^\text{I})^*\!\!\qquad \gamma^\text{II}=(\gamma^\text{I})^*\!\!\qquad \eta^\text{II}=(\eta^\text{I})^*
\end{array}
\end{equation}
where $\tan r_{\Omega}=e^{-\pi E_\Omega}$, $E_\Omega$ is the energy of the Rindler mode $k$, $E_\omega$ is the energy of the Minkowski mode $j$ and $\theta$ is a parameter defined such that it satisfies the condition $E_\Omega=m\cosh\theta$ and $|\bm k_\Omega|=m\sinh\theta$   (see \cite{Jauregui}). 
Finally, taking into account that $c_{j,\text{M}}=(u^+_{j,\text{M}},\phi)$ and $d^\dagger_{j,\text{M}}=(u^{-}_{j,\text{M}},\phi)$,  we find 
\begin{align}
\nonumber c_{j,\text{M}}=\sum_{k}\left[\alpha^{\text{I}*}_{jk}c_{k,\text{I}} + \beta^\text{I}_{jk} d_{k,\text{I}}^\dagger+\alpha^{\text{II}*}_{jk}c_{k,\text{II}}+\beta^\text{II}_{jk}d_{k,\text{II}}^{\dagger}\right],\\
d^\dagger_{j,\text{M}}=\sum_{k}\left[\gamma^{\text{I}*}_{jk}c_{k,\text{I}} + \eta^\text{I}_{jk} d_{k,\text{I}}^\dagger+\gamma^{\text{II}*}_{jk}c_{k,\text{II}}+\eta^\text{II}_{jk}d_{k,\text{II}}^{\dagger}\right].
\end{align}

We now consider the transformations between states in different basis. For this we define an arbitrary element of the Dirac field Fock basis for each mode as
\begin{equation} \label{tensprod}
\ket{F_k}=\ket{F_k}_\text{\text{R}}\otimes \ket{F_k}_\text{\text{L}},
\end{equation}
where 
\begin{eqnarray}
\nonumber\ket{F_k}_\text{R}&=&\ket{n}^+_\text{I}\ket{m}^-_\text{II},\\*
\ket{F_k}_\text{L}&=&\ket{p}^-_\text{I}\ket{q}^+_\text{II}.
\end{eqnarray}
Here the $\pm$ signs  denote particle/antiparticle. It is now convenient to introduce a new basis for inertial observers which corresponds to a superposition of Minkowski monocromatic modes.  The reason for this is that the new modes, called Unruh modes \cite{Jorma}, and Rindler modes have a simple Bogoliubov transformation: each Unruh mode transforms to a single frequency Rindler mode.  This transformation is given by
\begin{align}\label{Unruhop}
\nonumber C_{k,\text{\text{R}}}=&\left(\cos r_k\, c_{k,\text{I}}-\sin r_k\, d^\dagger_{k,\text{II}}\right),\\*
\nonumber C_{k,\text{\text{L}}}=&\left(\cos r_k\, c_{k,\text{II}}-\sin r_k\, d^\dagger_{k,\text{I}}\right),\\*
\nonumber D^\dagger_{k,\text{\text{R}}}=&\left(\sin r_k\, c_{k,\text{I}}+\cos r_k\, d^\dagger_{k,\text{II}}\right),\\*
D^\dagger_{k,\text{\text{L}}}=&\left(\sin r_k\, c_{k,\text{II}}+\cos r_k\, d^\dagger_{k,\text{I}}\right),
\end{align}
were $ C_{k,\text{\text{R,L}}}$ and $ D_{k,\text{\text{R,L}}}$ are the Unruh mode operators. 

The corresponding transformation between Minkowski and Unruh modes are given by,
\begin{align}
c_{j,\text{M}}=&N_j\sum_k\Bigg[\chi^*(C_{k,\text{\text{R}}}\otimes\openone_\text{\text{L}})+\chi(\openone_\text{\text{R}}\otimes C_{k,\text{\text{L}}})\Bigg],\\
\label{annihil} d^\dagger_{j,\text{M}}=&N_j\sum_k\Bigg[\chi(D^\dagger_{k,\text{\text{R}}}\otimes\openone_\text{\text{L}})+\chi^*(\openone_\text{\text{R}}\otimes D^\dagger_{k,\text{\text{L}}})\Bigg],
\end{align}
where
\begin{equation}\label{ene}
N_j=\frac{1}{2\sqrt{\pi E_\omega}}\qquad  \chi=(1+i)e^{i\theta E_{\Omega}}.
\end{equation}
Here we have written the tensor product structure \eqref{tensprod} explicitly.

 For massless fields it can be shown \cite{ch1} that  the Unruh operators have the same form as Eq. \eqref{Unruhop} however in this case $\tan r_{k}=e^{-\pi\Omega/a}$.

In the massless case, to find the Minkowski vacuum in the Rindler basis we consider the following ansatz
\begin{equation}
\ket{0}_\text{M}=\bigotimes_{\Omega}\ket{0_\Omega}_\text{M},
\end{equation}
where $\ket{0_\Omega}_\text{M}=\ket{0_\Omega}_\text{\text{R}}\otimes\ket{0_\Omega}_\text{\text{L}}$.
We find that
\begin{eqnarray}
\label{vauno}
\ket{0_{\Omega}}_\text{\text{R}}&=&\sum_{n,s}\left(F_{n,\Omega,s}\ket{n_{\Omega,s}}^+_\text{I}\ket{n_{\Omega,-s}}^-_\text{II}\right)\nonumber \\
\ket{0_{\Omega}}_\text{\text{L}}&=&\sum_{n,s}\left(G_{n,\Omega,s}\ket{n_{\Omega,s}}^-_\text{I}\ket{n_{\Omega,-s}}^+_\text{II}\right),
\end{eqnarray}
where the label $\pm$ denotes particle/antiparticle modes  and $s$ labels the spin. The minus signs on the spin label in region $\text{II}$ show explicitly that spin, as all the magnitudes which change under time reversal, is opposite in region I with respect to region II.

We obtain the form of the coefficients $F_{n,\Omega,s},G_{n,\Omega,s}$ for the vacuum by imposing that the Minkowski vacuum is annihilated by the particle annihilator for all frequencies and values for the spin third component.

Since the simplest case that preserves the fundamental Dirac characteristics corresponds to Grassman scalars, we study them in what follows. In this case, the Pauli exclusion principle limits the sums \eqref{vauno} and only the two following terms contribute
\begin{eqnarray}
\label{vaunos}
\ket{0_\Omega}_\text{\text{R}}&=&F_0\ket{0_\Omega}^+_\text{I}\ket{0_\Omega}^-_\text{II}+F_1\ket{1_\Omega}^+_\text{I}\ket{1_\Omega}^-_\text{II},\nonumber \\
\ket{0_\Omega}_\text{\text{L}}&=&G_0\ket{0_\Omega}^-_\text{I}\ket{0_\Omega}^+_\text{II}+G_1\ket{1_\Omega}^-_\text{I}\ket{1_\Omega}^+_\text{II}.
\end{eqnarray}
Due to the anticommutation relations we must introduce the following sign conventions
\begin{eqnarray}
\ket{1_\Omega}^+_\text{I}\!\ket{1_\Omega}^-_\text{II}&=&c^\dagger_{\Omega,\text{I}}d^\dagger_{\Omega,\text{II}}\!\ket{0_\Omega}^+_\text{I}\!\ket{0_\Omega}^-_\text{II},\nonumber \\
&=&-d^\dagger_{\Omega,\text{II}}c^\dagger_{\Omega,\text{I}}\!\ket{0_\Omega}^+_\text{I}\!\ket{0_\Omega}^-_\text{II},\nonumber\\*
\nonumber\ket{1_\Omega}^-_\text{I}\!\ket{1_\Omega}^+_\text{II}&=&d^\dagger_{\Omega,\text{I}}c^\dagger_{\Omega,\text{II}}\!\ket{0_\Omega}^-_\text{I}\!\ket{0_\Omega}^+_\text{II},\nonumber \\
&=&-c^\dagger_{\Omega,\text{I}V}d^\dagger_{\Omega,\text{I}}\!\ket{0_\Omega}^-_\text{I}\!\ket{0_\Omega}^+_\text{II}.
\end{eqnarray}
For the case of Grassman scalars  \cite{Jorma}, after imposing that $c_{\omega,\text{M}}\ket{0_\Omega}_\text{M}=0$ for all $\omega$ omega we obtain $C_{\Omega,\text{\text{R}}}\ket{0_\Omega}_\text{\text{R}}=C_{\Omega,\text{\text{L}}}\ket{0_\Omega}_\text{\text{L}}=0$ for all $\Omega$.  The vacuum state then yields
 \begin{align}\label{vacgrassman} 
\nonumber\ket{0_{\Omega}}&=\left(\cos r_{\Omega}\ket{0_{\Omega}}^+_\text{I}\ket{0_{\Omega}}^-_\text{II}+\sin r_{\Omega}\ket{1_{\Omega}}^+_\text{I}\ket{1_{\Omega}}^-_\text{II}\right)\\
&\otimes\left(\cos r_{\Omega}\ket{0_{\Omega}}^-_\text{I}\ket{0_{\Omega}}^+_\text{II}-\sin r_{\Omega}\ket{1_{\Omega}}^-_\text{I}\ket{1_{\Omega}}^+_\text{II}\right)\!,
\end{align}
Using equation \eqref{annihil} that this vacuum state also satisfied $d_{\omega,\text{M}}\ket{0_\Omega}_\text{M}=0\; \forall\omega$ which is equivalent to $D_{\Omega,\text{\text{R}}}\ket{0_\Omega}_\text{\text{R}}=D_{\Omega,\text{\text{L}}}\ket{0_\Omega}_\text{\text{L}}=0\;\forall\Omega$.
For convenience, we introduce the following compact notation, \begin{equation}\label{shortnot}
\ket{i j k l}_{\Omega}\equiv\ket{i_{\Omega}}^+_\text{I}\ket{j_{\Omega}}^-_\text{I}\ket{k_{\Omega}}^+_\text{II}\ket{l_{\Omega}}^-_\text{II},
\end{equation}
which is slightly different from the one employed in \cite{Jorma}. In this notation the vacuum state is written as,
\begin{eqnarray}\label{shortvac}
\ket{0_{\Omega}}&=&\cos^2 r_{\Omega}\ket{0000}_{\Omega}-\sin r_{j}\cos r_{\Omega}\ket{0110}_{\Omega}\\
&+&\sin r_{\Omega}\cos r_{\Omega} \ket{1001}_{\Omega}-\sin^2 r_{\Omega} \ket{1111}_{\Omega}.\nonumber
\end{eqnarray}

The Minkowskian one particle state is obtained by applying the creation operator of particle or antiparticle to the vacuum state $\ket{1_{j}}^+_{\text{U}}=c_{\Omega,\text{U}}^\dagger\ket{0}_\text{M}$, $\ket{1_{j}}^-_{\text{U}}=d_{\Omega,\text{U}}^\dagger\ket{0}_\text{M}$
where the Unruh particle/antiparticle creator is a combination of the two Unruh operators 
\begin{align}\label{creat}
%Correcci—n por omegas las k de anterior versi—n
c_{\Omega,\text{U}}^\dagger=&q_\text{\text{R}}(C^\dagger_{\Omega,\text{\text{R}}}\otimes\openone_\text{\text{L}})+q_\text{\text{L}}(\openone_\text{\text{R}}\otimes C^\dagger_{\Omega,\text{\text{L}}}),\nonumber\\
d_{\Omega,\text{U}}^\dagger=&p_\text{\text{R}}(D^\dagger_{\Omega,\text{\text{R}}}\otimes\openone_\text{\text{L}})+p_\text{\text{L}}(\openone_\text{\text{R}}\otimes D^\dagger_{\Omega,\text{\text{L}}}).
\end{align}
$q_\text{\text{R}},q_\text{\text{L}},p_\text{\text{R}},p_\text{\text{L}}$ are complex numbers satisfying $|q_\text{\text{R}}|^2+|q_\text{\text{L}}|^2=1$,  $|p_\text{\text{R}}|^2+|p_\text{\text{L}}|^2=1$.

The parameters $p_\text{R,L}$  are not independent of $q_\text{R,L}$. We demand that the Unruh particle and antiparticle operators are referred to particle and antiparticle modes in the same Rindler wedges. Therefore to be coherent with a particular election of $q_\text{R}$ and $q_\text{L}$,  we have to choose $p_\text{L}=q_\text{R}$ and $p_\text{R}=q_\text{L}$, 
\begin{align}\label{creat}
c_{k,\text{U}}^\dagger=&q_\text{\text{R}}(C^\dagger_{\Omega,\text{\text{R}}}\otimes\openone_\text{\text{L}})+q_\text{\text{L}}(\openone_\text{\text{R}}\otimes C^\dagger_{\Omega,\text{\text{L}}}),\nonumber\\
d_{k,\text{U}}^\dagger=&q_\text{\text{L}}(D^\dagger_{\Omega,\text{\text{R}}}\otimes\openone_\text{\text{L}})+q_\text{\text{R}}(\openone_\text{\text{R}}\otimes D^\dagger_{\Omega,\text{\text{L}}}).
\end{align}

The Unruh L and R field excitations are given by
\begin{align}
\nonumber\ket{1_{\Omega}}^+_\text{\text{R}}=&C^\dagger_{\Omega,\text{\text{R}}}\ket{0_{\Omega}}_\text{\text{R}}=\ket{1_{\Omega}}^+_\text{I}\ket{0_{\Omega}}^-_\text{II}\\*
\nonumber \ket{1_{\Omega}}^+_\text{\text{L}}=&C^\dagger_{\Omega,\text{\text{L}}}\ket{0_{\Omega}}_\text{\text{L}}=\ket{0_{\Omega}}^-_\text{I}\ket{1_{\Omega}}^+_\text{II}\\
\nonumber\ket{1_{\Omega}}^-_\text{\text{R}}=&D^\dagger_{\Omega,\text{\text{R}}}\ket{0_{\Omega}}_\text{\text{R}}=\ket{0_{\Omega}}^+_\text{I}\ket{1_{\Omega}}^-_\text{II}\\*
\ket{1_{\Omega}}^-_\text{\text{L}}=&D^\dagger_{\Omega,\text{\text{L}}}\ket{0_{\Omega}}_\text{\text{L}}=\ket{1_{\Omega}}^-_\text{I}\ket{0_{\Omega}}^+_\text{II}
\end{align}
and therefore,
\begin{align}
\nonumber \ket{1_{k}}^+_\text{U}=&q_\text{\text{R}}\ket{1_{\Omega}}^+_\text{\text{R}}\otimes\ket{0_{\Omega}}_\text{\text{L}}+q_\text{\text{L}}\ket{0_{\Omega}}_\text{\text{R}}\otimes\ket{1_{\Omega}}^+_\text{\text{L}},\\
\ket{1_{k}}^-_\text{U}=&q_\text{\text{L}}\ket{1_{\Omega}}^-_\text{\text{R}}\otimes\ket{0_{\Omega}}_\text{\text{L}}+q_\text{\text{R}}\ket{0_{\Omega}}_\text{\text{R}}\otimes\ket{1_{\Omega}}^-_\text{\text{L}}.
\end{align}
In the short notation we have introduced the state reads,  
\begin{eqnarray}\label{onegrassman}
\nonumber\ket{1_{k}}^+_\text{U}&=& q_\text{\text{R}}\left[\cos r_{k}\ket{1000}_{\Omega}-\sin r_{\Omega}\ket{1110}_{\Omega}\right]\\*
&+&q_\text{\text{L}}\left[\cos r_{\Omega}\ket{0010}_{\Omega}+\sin r_{\Omega}\ket{1011}_{\Omega}\right],\nonumber\\[4mm]
\nonumber\ket{1_{k}}^-_\text{U}&=& q_\text{\text{L}}\left[\cos r_{k}\ket{0001}_{\Omega}-\sin r_{\Omega}\ket{0111}_{\Omega}\right]\\*
&+&q_\text{\text{R}}\left[\cos r_{\Omega}\ket{0100}_{\Omega}+\sin r_{\Omega}\ket{1101}_{\Omega}\right].
\end{eqnarray}

\section{Particle and Anti-particle entanglement in non-inertial frames}\label{sec2}

Having the expressions for the vacuum and single particle states in the  Unruh and Rindler bases enables us to analyse the degradation of entanglement from the perspective of observers in uniform acceleration.  Let us consider the following maximally entangled states from the inertial perspective

\begin{align}
\label{1e}\ket{\Psi_+}=&\frac{1}{\sqrt2}\left(\ket{0_\omega}_{\text M}\ket{0_\Omega}_\text{U}+\ket{1_\omega}^\sigma_{\text{M}}\ket{1_\Omega}^+_\text{U}\right),\\
\label{2e}\ket{\Psi_-}=&\frac{1}{\sqrt2}\left(\ket{0_\omega}_{\text M}\ket{0_\Omega}_\text{U}+\ket{1_\omega}^\sigma_{\text{M}}\ket{1_\Omega}^-_\text{U}\right),\\
\label{3e}\ket{\Psi_1}=&\frac{1}{\sqrt2}\left(\ket{1_\omega}^+_{\text M}\ket{1_\Omega}^-_\text{U}+\ket{1_\omega}^-_{\text{M}}\ket{1_\Omega}^+_\text{U}\right),
\end{align}
where the modes labeled with $\text{U}$ are Grassman Unruh modes  and the label $\sigma=\pm$ denotes particle and antiparticle modes. The first two states correspond to entangled states with particle and antiparticle Unruh excitations, respectively. These two states are analogous to the bosonic state $\frac{1}{\sqrt2}(\ket{0}_M\ket{0}_U+\ket{1}_M\ket{1}_U)$ which is entangled in the  occupation number degree of freedom.  The third state has no analog in the neutral bosonic scenario since in this case the state is entangled in the particle/antiparticle degree of freedom. In spite that fermionic entanglement in non-inertial frames  has been extensively studied in the literature \cite{AlsingSchul},  states \eqref{2e} and \eqref{3e} have not been considered before.

We consider Alice to be an inertial observer with a detector sensitive to $\omega$ modes while her partner Rob who is in uniform acceleration carries a detector sensitive to $\Omega$ modes.  To  study the entanglement in the states  from their perspective we must transform the $\Omega$ modes to Rindler modes.  Therefore, Unruh states must be transformed into the Rindler basis. The state in the Minkowski-Rinder basis becomes effectively  a tri-partite system.  As it is commonplace in the literature,  we define the Alice-Rob bi-partition as the Minkowski and region I Rindler modes while the Alice-AntiRob bi-partitions corresponds to Minkowski and region II Rindler modes.  To study distillable entanglement we will employ the negativity $\mathcal{N}$, defined as the sum of the negative eigenvalues of the partial transpose density matrix. Two cases of interest will be considered. In the first case we assume that Alice and Rob have detectors which do not distinguish between particle and antiparticles. In this case,  particles and antiparticles together are considered to be a subsystem. In the second case we consider that Rob and AntiRob have detectors which are only sensitive to particles (antiparticles) therefore, antiparticle (particle) states must be traced out. Our results will show that when Rob is accelerated, the entanglement redistributes between particles and antiparticles as a function of his acceleration. This effect is a unique feature of fermionic fields and plays an important role in the behavior of fermionic  entanglement in the infinite acceleration limit.

\subsection{Entanglement in states $\ket{\Psi_+}$ and $\ket{\Psi_-}$ }

To compute Alice-Rob partial density matrix  in \eqref{1e}  we trace over region II in $\proj{\Psi_+}{\Psi_+}$ and obtain,  
\begin{eqnarray}\label{ARd}
\nonumber\rho^+_{AR}\!\!&=&\!\!\frac{1}{2}\Big[C^4\proj{000}{000}+S^2C^2(\proj{010}{010}+\proj{001}{001})\\
\nonumber&&\!\!\!\!\!\!\!\!\!\!\!\!\!\!\!\!\!+\phantom{\Big[}S^4\proj{011}{011}+|q_\text{\text{R}}|^2(C^2\proj{110}{110}+S^2\proj{111}{111})\\*
\nonumber&&\!\!\!\!\!\!\!\!\!\!\!\!\!\!\!\!\!+\phantom{\Big[}\!\!|q_\text{\text{L}}|^2(S^2\!\proj{110}{110}\!+\!C^2\!\proj{100}{100})\!+\!q_\text{\text{R}}^*(C^3\!\proj{000}{110}\\*
\nonumber&&\!\!\!\!\!\!\!\!\!\!\!\!\!\!\!\!\!+S^2C\!\proj{001}{111})-q_\text{\text{L}}^*(C^2S\!\proj{001}{100}+S^3\!\proj{011}{110})\\*
&&\!\!\!\!\!\!\!\!\!\!\!\!\!\!\!\!\!-q_\text{R} q_\text{L}^*SC\!\proj{111}{100}\Big]+(\text{H.c.})_{_{\substack{\text{non-}\text{diag.}}}}
\end{eqnarray}
were $C=\cos r_{\Omega}$ and $S=\sin r_{\Omega}$. 

The density matrix for the Alice-AntiRob modes is obtained by tracing over region $\text{I}$, \begin{eqnarray}\label{AaRd}
\nonumber\rho^+_{A\bar R}\!\!&=&\!\!\frac{1}{2}\Big[C^4\proj{000}{000}+S^2C^2(\proj{001}{001}+\proj{010}{010})\\*
\nonumber&&\!\!\!\!\!\!\!\!\!\!\!\!\!\!\!\!\!+\phantom{\Big[}S^4\proj{011}{011}+|q_\text{\text{R}}|^2(C^2\proj{100}{100}+S^2\proj{110}{110})\\*
\nonumber&&\!\!\!\!\!\!\!\!\!\!\!\!\!\!\!\!\!+\!\!\phantom{\Big[}|q_\text{\text{L}}|^2(S^2\!\proj{111}{111}\!+\!C^2\!\proj{110}{110})\!+\!q_\text{\text{L}}^*(C^3\!\proj{000}{110}\\*
\nonumber&&\!\!\!\!\!\!\!\!\!\!\!\!\!\!\!\!\!+S^2C\!\proj{001}{111})+q_\text{\text{R}}^*(C^2S\!\proj{001}{100}+S^3\!\proj{011}{110})\\*
&&\!\!\!\!\!\!\!\!\!\!\!\!\!\!\!\!\!+q_\text{R} q_\text{L}^*SC\!\proj{100}{111}\Big]+(\text{H.c.})_{_{\substack{\text{non-}\\\text{diag.}}}}
\end{eqnarray}

To calculate the entanglement considering that Rob and AntiRob are able to detect both particles and antiparticles (the calculations follows from  \cite{Jorma}), 
we first obtain the Alice-Rob and Alice-AntiRob partial transpose density matrices  and their negative eigenvalues. The partial transpose matrix $(\rho^+_{AR})^{pT}$ is block diagonal and only the following two blocks contribute to negativity,
\begin{itemize}
\item Basis: $\{\ket{100},\ket{010},\ket{111}\}$
\end{itemize}
\begin{equation}
\frac12\left(\begin{array}{ccc}
 C^2|q_\text{\text{L}}|^2& C^3q_\text{\text{R}}^*&-q_\text{R}^* q_\text{L}SC\\
C^3q_\text{\text{R}}& S^2C^2& -q_\text{\text{L}} S^3\\
-q_\text{R} q_\text{L}^*SC & -q_\text{\text{L}}^* S^3&|q_\text{\text{R}}|^2S^2\\
\end{array}\right),
\end{equation}
\begin{itemize}
\item Basis: $\{\ket{000},\ket{101},\ket{011}\}$
\end{itemize}
\begin{equation}
\frac12\left(\begin{array}{ccc}
C^4&-q_\text{\text{L}}C^2S&0\\
-q_\text{\text{L}}^*C^2S& 0 & q_\text{\text{R}}^*S^2C \\
0& q_\text{\text{R}}S^2C & S^4 \\
\end{array}\right).
\end{equation}
were the basis used is $\ket{ijk}=\ket{i}_\text{M}^\sigma\overbrace{\ket{j}_\text{I}^+\ket{k}^-_\text{I}}^{\text{Rob}}$.  Notice that although the system is bipartite, the dimension of the partial Hilbert space for Alice is smaller than the dimension of the Hilbert space for Rob, which includes both particle and antiparticle modes.
The eigenvalues only depend on $|q_\text{\text{R}}|$ and not on the relative phase between $q_\text{\text{R}}$ and $q_\text{\text{L}}$.

We can carry out a similar calculation for the Alice-AntiRob subsystem. In this case we need to compute $(\rho^+_{A\bar R})^{pT}$  the blocks of the partial transpose density matrix which contribute to the negativity are, 
\begin{itemize}
\item Basis: $\{\ket{111},\ket{010},\ket{100}\}$
\end{itemize}
\begin{equation}
\frac12\left(\begin{array}{ccc}
 S^2|q_\text{\text{L}}|^2& S^3q_\text{\text{R}}^*&q_\text{R}^* q_\text{L}SC\\
S^3q_\text{\text{R}}& C^2S^2& q_\text{\text{L}} C^3\\
q_\text{R} q_\text{L}^*SC & q_\text{\text{L}}^* C^3&|q_\text{\text{R}}|^2C^2\\
\end{array}\right),
\end{equation}
\begin{itemize}
\item Basis: $\{\ket{011},\ket{101},\ket{000}\}$
\end{itemize}
\begin{equation}
\frac12\left(\begin{array}{ccc}
S^4&q_\text{\text{L}}S^2C&0\\
q_\text{\text{L}}^*S^2C& 0 & q_\text{\text{R}}^*C^2S \\
0& q_\text{\text{R}}C^2S & C^4 \\
\end{array}\right),
\end{equation}
where we have considered the basis $\ket{ijk}=\ket{i}_\text{M}^\sigma\overbrace{\ket{j}_\text{II}^+\ket{k}^-_\text{II}}^{\text{Anti-Rob}}$. Once more, the eigenvalues only depend on $|q_\text{\text{R}}|$ and not on the relative phase between $q_\text{\text{R}}$ and $q_\text{\text{L}}$.

\begin{figure}[h]
\begin{center}
\includegraphics[width=.50\textwidth]{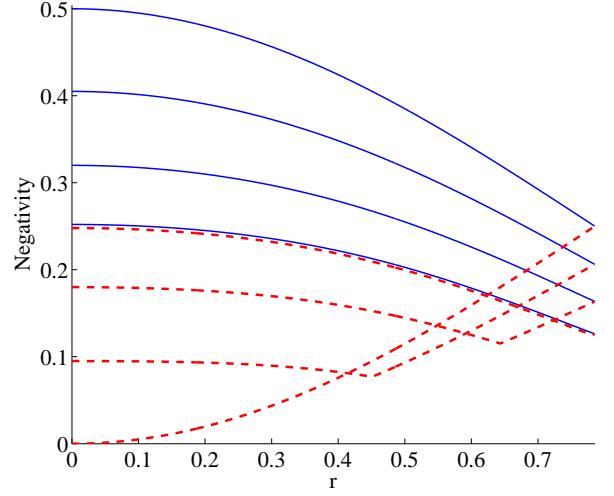}
\end{center}
\caption{Full Negativity for the Alice-Rob (Blue continuous) and Alice-AntiRob (red dashed) bipartitions as a function of $r_{\Omega}=\arctan{e^{-\pi\Omega/a}}$ for various choices of $|q_\text{\text{R}}|$. The blue continuous (red dashed) curves from top to bottom (from bottom to top) correspond, to $|q_\text{\text{R}}|=1,0.9,0.8,0.7$  respectively.}
\label{N0}
\end{figure}

We see that, as discussed in \cite{Jorma}, when the acceleration increases the entanglement between Alice-AntiRob modes is created compensating the entanglement lost between Alice-Rob. If $|q_R|<1$ the entanglement lost is not entirely compensated by the creation of entanglement between Alice-AntiRob resulting in a less entangled state in the infinite acceleration limit. An analysis of the quantum entanglement between Alice's modes and particle/antiparticle modes of Rob and AntiRob will be useful to disclose why correlations present this behaviour. 

We now analyse the entanglement when Rob's and AntiRob's detectors are not able to detect antiparticles.  In this case the entanglement  is between their particle modes and Alice's subsystem. Since Rob cannot detect antiparticles we must trace over all antiparticle states and therefore, \eqref{ARd}: $\rho^+_{AR^+}=\sum_{n=0,1}\bra{n}_{\text{I}}^-\rho^+_{AR}\ket{n}_{\text{I}}^-$. This yields
\begin{eqnarray}
\nonumber\rho^+_{AR^+}&=&\frac{1}{2}\Big[C^2\proj{00}{00}+S^2\proj{01}{01}+q_\text{\text{R}}^*C\!\proj{00}{11})\\
\nonumber&+&(|q_\text{\text{R}}|^2+|q_\text{\text{L}}|^2S^2)\proj{11}{11}+|q_\text{\text{L}}|^2C^2\!\proj{10}{10})\Big]\\*
&+&(\text{H.c.})_{_{\substack{\text{non-}\text{diag.}}}}
\end{eqnarray}
which is the partial state of Alice and the particles sector of Rob.

The partial transpose $(\rho^+_{AR^+})^{pT}$ has only one block whose negative eigenvalue contributes to negativity
\begin{itemize}
\item Basis: $\{\ket{10},\ket{01}\}$
\end{itemize}
\begin{equation}
\frac12\left(\begin{array}{cc}
|q_\text{\text{L}}|^2C^2&q_\text{\text{R}}^*C\\
q_\text{\text{R}}C& S^2  
\end{array}\right),
\end{equation}

The same procedure can be carried out for the system $A\bar R$ tracing over the antiparticle sector in \eqref{AaRd} obtaining 
\begin{eqnarray}
\nonumber\rho^+_{A\bar R^+}&=&\frac{1}{2}\Big[C^2\proj{00}{00}+S^2\proj{01}{01}+q_\text{\text{L}}^*C\proj{00}{11}\\*
\nonumber&+&(|q_\text{L}|^2+|q_\text{R}|^2S^2)\proj{11}{11}+|q_\text{\text{R}}|^2C^2\proj{10}{10}\Big]\\*
&+&(\text{H.c.})_{_{\substack{\text{non-}\\\text{diag.}}}}
\end{eqnarray}
The partial transpose $(\rho^+_{A\bar R^+})^{pT}$ has only one block whose negative eigenvalue contributes to negativity
\begin{itemize}
\item Basis: $\{\ket{10},\ket{01}\}$
\end{itemize}
\begin{equation}
\frac12\left(\begin{array}{cc}
|q_\text{\text{R}}|^2C^2&q_\text{\text{L}}^*C\\
q_\text{\text{L}}C& S^2  
\end{array}\right),
\end{equation}

\begin{figure}[h]
\begin{center}
\includegraphics[width=.50\textwidth]{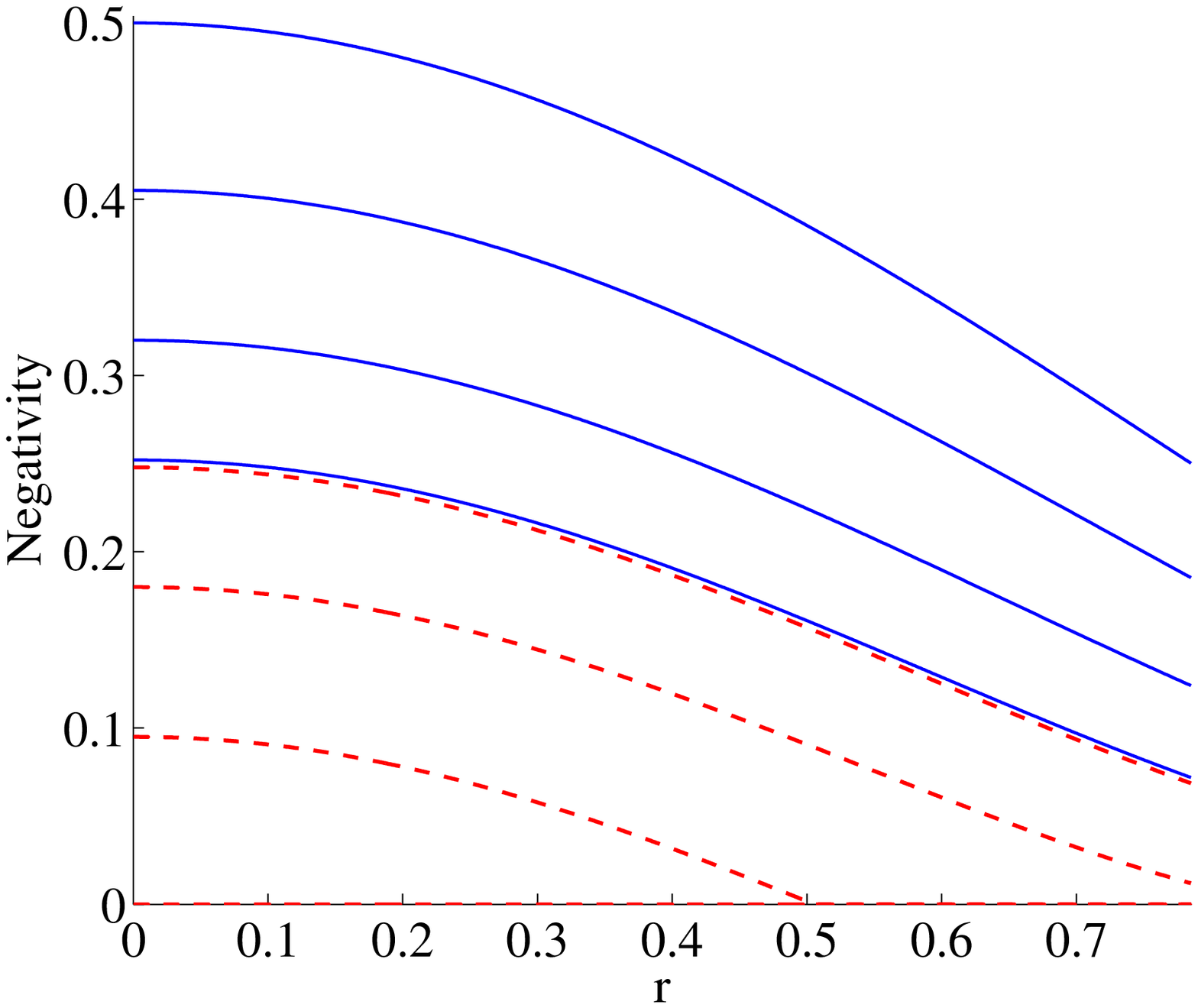}
\end{center}
\caption{Negativity in the particle sector (Rob and AntiRob can only detect particle modes) for the bipartition Alice-Rob (Blue continuous) and Alice-AntiRob (red dashed) for the state \eqref{1e} as a function of $r_{\Omega}=\arctan{e^{-\pi\Omega/a}}$ for various choices of $|q_\text{\text{R}}|$. The blue continuous (red dashed) curves from top to bottom (from bottom to top) correspond, to $|q_\text{\text{R}}|=1,0.9,0.8,0.71$  respectively. For $|q_R|=1$ Alice-AntiRob curve is zero $\forall a$}
\label{Npp}
\end{figure}

A similar calculation can be carried out considering that Rob and AntiRob detectors are only sensitive to antiparticles, i.e. tracing over particle states. In this case we obtain,
\eqref{ARd}: $\rho^+_{AR^-}=\sum_{n=0,1}\bra{n}_{\text{I}}^+\rho^+_{AR}\ket{n}_{\text{I}}^+$, and therefore, 
\begin{eqnarray}
\nonumber\rho^+_{AR^-}&=&\frac{1}{2}\Big[C^2\proj{00}{00}+S^2\proj{01}{01}-q_\text{\text{L}}^*S\proj{01}{10}\\
\nonumber&+&(|q_\text{\text{L}}|^2+|q_\text{\text{R}}|^2 C^2)\proj{10}{10}+|q_\text{\text{R}}|^2 S^2\proj{11}{11})\Big]\\*
&+&(\text{H.c.})_{_{\substack{\text{non-}\text{diag.}}}}
\end{eqnarray}
is the partial state of Alice and the particles sector of Rob.

The only block giving negative eigenvalues is
\begin{itemize}
\item Basis: $\{\ket{11},\ket{00}\}$
\end{itemize}
\begin{equation}
\frac12\left(\begin{array}{cc}
|q_\text{\text{R}}|^2S^2&-q_\text{\text{L}}^*S\\
-q_\text{\text{L}}S& C^2  
\end{array}\right),
\end{equation}

The density matrix for the Alice-AntiRob antiparticle modes is obtained by tracing over the particle sector in \eqref{AaRd}
\begin{eqnarray}
\nonumber\rho^+_{A\bar R^-}&=&\frac{1}{2}\Big[C^2\proj{00}{00}+S^2\proj{01}{01}\\*
\nonumber&+&(|q_\text{\text{R}}|^2+|q_\text{\text{L}}|^2C^2)\proj{10}{10}+|q_\text{\text{L}}|^2S^2\!\proj{11}{11}\\*
&+&q_\text{\text{R}}^*S\proj{01}{10})\Big]+(\text{H.c.})_{_{\substack{\text{non-}\\\text{diag.}}}}
\end{eqnarray}

Again only one block of the density matrix contributes to negativity
\begin{itemize}
\item Basis: $\{\ket{11},\ket{00}\}$
\end{itemize}
\begin{equation}
\frac12\left(\begin{array}{cc}
|q_\text{\text{L}}|^2S^2&q_\text{\text{R}}^*S\\
q_\text{\text{R}}S& C^2  
\end{array}\right),
\end{equation}

\begin{figure}[h]
\begin{center}
\includegraphics[width=.50\textwidth]{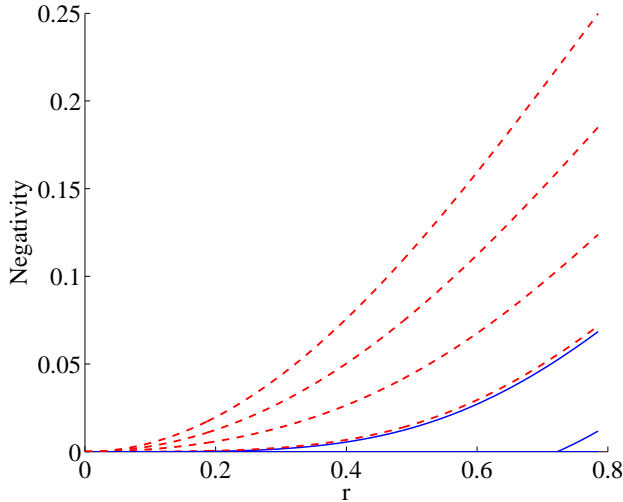}
\end{center}
\caption{Negativity in the antiparticle sector (Rob and AntiRob can only detect antiparticle modes) for the bipartition Alice-Rob (Blue continuous) and Alice-AntiRob (red dashed) for the state \eqref{1e} as a function of $r_{\Omega}=\arctan{e^{-\pi\Omega/a}}$ for various choices of $|q_\text{\text{R}}|$. The red dashed (blue continuous) curves from top to bottom (from bottom to top) correspond to $|q_\text{\text{R}}|=1,0.9,0.8,0.71$  respectively.}
\label{Npm}
\end{figure}

The analysis of entanglement in the state \eqref{2e} is done in a completely analogous way. We find that the entanglement behaves exactly the same way as in state \eqref{1e} only that the role of particles is replaced by anti-particles. Therefore negativities  are related in the following way
\[\mathcal{N}_{AR^+}^+=\mathcal{N}_{A R^-}^-\quad\mathcal{N}_{A\bar R^+}^+=\mathcal{N}_{A\bar R^-}^-\]
\[\mathcal{N}_{AR^-}^+=\mathcal{N}_{A R^+}^-\quad\mathcal{N}_{A\bar R^-}^+=\mathcal{N}_{A\bar R^+}^-\]
\[\mathcal{N}_{AR}^+=\mathcal{N}_{A R}^-\quad\mathcal{N}_{A\bar R}^+=\mathcal{N}_{A \bar R}^-\]

We see in figure \ref{N0} that the total entanglement between Alice and AntiRob starts decreasing and presents a minimum before starting to grow again for higher accelerations. If $|q_R|<1$ the entanglement in the limit $a\rightarrow0$ is distributed between the bipartitions $AR$ and $A\bar R$ \cite{Jorma}. The entanglement lost in the bipartition  $A\bar R$ is not entirely compensated by the creation of entanglement in  $A\bar R$  and therefore, this results in a state containing less entanglement in the infinite acceleration limit. 

Interestingly, the correlations between Alice and the particle sector of Rob and AntiRob always decrease (Fig. \ref{Npp}) while the correlations between Alice and the antiparticle sector of Rob and AntiRob always grow (Fig. \ref{Npm}). This behaviour explains why entanglement always survive the infinite acceleration limit for any election of $q_R$ and $q_L$. As Rob accelerates there is a process of entanglement transfer between the particle and antiparticle sector of his Hilbert space. The same happens with AntiRob, such that neither for $AR$ nor $A\bar R$ the entanglement vanishes for any value of the acceleration.

For the simplest case $|q_R|=1$ we see that all the entanglement is initially ($a\rightarrow0$) in the particle sector of the bipartition $AR$. As the acceleration increases the entanglement is transferred to the antiparticle sector of the bipartition $A\bar R$ such that, in the limit of infinite acceleration entanglement has been equally distributed between these two bipartitions.  

The tensor product structure of the particle and antiparticle sectors \eqref{vacgrassman} plays an important role in the behavior of entanglement in the infinite acceleration limit.  In the case of neutral scalar fields there are no antiparticles and entanglement is completely degraded. Note that in the case of charged bosonic fields there are indeed charged conjugate antiparticles. However, in this case the Hilbert space has a similar structure to the uncharged field  \cite{Alicefalls}. The existence of  bosonic antiparticles simply adds another copy of the same Hilbert space and no entanglement transfer is possible between particle and antiparticles. Clearly, the Hilbert space structure in the fermionic case \eqref{vacgrassman} is different and thus, we observe the differences in the entanglement behavior.

Now, if we move to less trivial cases where $q_R\neq 1$ the situation gets more complicated. In these scenarios we initially start with some entanglement in the particle sector of $A\bar R$, and there can also be an entanglement transfer between this sector and $AR$ antiparticle sector. However  no entanglement at all is transferred to the antiparticle sector of the subsystem $AR$ unless the acceleration reaches a threshold given by
\begin{equation}\label{thres}
  \cos^2r =\frac{|q_L|^2}{|q_R|^2}=\frac{1}{|q_R|^2}-1.
  \end{equation}
The maximum value of $\cos \left[r(a)\right]$ is $\cos \left[r(a\rightarrow\infty)\right]\rightarrow1/\sqrt2$  therefore for $|q_R|^2>2/3$  entanglement  is  not  transferred to the antiparticle sector of Alice-Rob for any value of the acceleration. This explains why when $q_R\neq1$ the entanglement loss in the bipartition AR in the limit $a\rightarrow\infty$ is smaller than in the extreme case $q_R=1$.
 
It is evident that the choice of Unruh modes influence the transfer of entanglement between particle and antiparticle sectors. When the acceleration is larger than \eqref{thres},  when $|q_\text{R}|$ grows closer  to $1/\sqrt{2}$ more entanglement is transferred from the particle sector of $A\bar R$ to the antiparticles of $AR$. In the limit $q_\text{R}=1/\sqrt2$ the same amount of entanglement  is transferred to the antiparticle sector of both $AR$ and $A \bar R$.

The particle entanglement (always monotonically decreasing) resembles the behaviour of bosonic entanglement studied in \cite{Jorma}. Bosonic entanglement is monotonically decreasing for both $AR$ and $A\bar R$ subsystems. In the bosonic case there are not antiparticles and, hence, there is no possibility of entanglement transfer to antiparticle sectors of $A\bar R$. This is the origin of the differences in entanglement behaviour for neutral bosons and fermions.

\subsection{Entanglement in state $\ket{\Psi_1}$}

This state has no neutral bosonic analog since it is entangled in the particle/antiparticle degree of freedom. Therefore, the analysis of entanglement in this state revels interesting features which are of genuinely fermionic nature.

To study this type of state we employ our generalization of the formalism developed in \cite{Jorma} which relates general Unruh modes with Rindler modes. This formalism refines the single-mode approximation introduced in \cite{Alsingtelep,AlsingMcmhMil}  which has been extensively used in the literature.  For this type of state the single approximation used in  \cite{AlsingSchul}  does not hold and attempting to use it leads to misleading results: one finds that maximally entangled states from the inertial perspective appear disentangled  from the accelerated perspective, irrespectively of the value of acceleration. Using the mode transformation introduced in our first section leads to sensible results: acceleration behaves regularly for accelerated observers and approaches a maximally entangled state in the inertial limit.

For convenience we will introduce a new notation for this case. For Alice, we denote the states by $\ket{+}$ if they correspond to particles and $\ket{-}$ for antiparticles. Therefore the state is written as
\begin{equation}
\ket{\Psi_1}=\frac{1}{\sqrt2}\left(\ket{+}_{\text M}\ket{1_\Omega}^-_\text{U}+\ket{-}_{\text{M}}\ket{1_\Omega}^+_\text{U}\right)
\end{equation}
The density matrix for the subsystem Alice-Rob is obtained from $\proj{\Psi_1}{\Psi_1}$  tracing over region II
\begin{align}
\nonumber\rho^1_{AR}=&|q_\text{R}|^2C^2\proj{+10}{+10}+|q_\text{R}|^2S^2\proj{+11}{+11}\\
\nonumber+&|q_\text{L}|^2C^2\proj{+00}{+00}+|q_\text{L}|^2S^2\proj{+10}{+10} \\
\nonumber-&SCq_\text{R}q_\text{L}^*\proj{+11}{+00}+|q_\text{L}|^2C^2\proj{-00}{-00}\\
\nonumber+&|q_\text{L}|^2S^2\proj{-01}{-01}+|q_\text{R}|^2C^2\proj{-01}{-01}\\
\nonumber+&|q_\text{R}|^2S^2\proj{-11}{-11}+SCq_\text{R}^*q_\text{L}\proj{-00}{-11}\\
+&(|q_\text{R}|^2C^2-|q_\text{L}|^2S^2)\proj{+10}{-01}+(\text{H.c.})_{_{\substack{\text{non-}\\\text{diag.}}}}
\end{align}
and for the Alice-AntiRob partition we obtain
\begin{align}
\nonumber\rho^1_{A\bar R}=&|q_\text{R}|^2C^2\proj{+00}{+00}+|q_\text{R}|^2S^2\proj{+10}{+10}\\
\nonumber+&|q_\text{L}|^2C^2\proj{+10}{+10}+|q_\text{L}|^2S^2\proj{+11}{+11} \\
\nonumber+&SCq_\text{R}q_\text{L}^*\proj{+00}{+11}+|q_\text{L}|^2C^2\proj{-01}{-01}\\
\nonumber+&|q_\text{L}|^2S^2\proj{-11}{-11}+|q_\text{R}|^2C^2\proj{-00}{-00}\\
\nonumber+&|q_\text{R}|^2S^2\proj{-01}{-01}-SCq_\text{R}^*q_\text{L}\proj{-11}{-00}\\
+&(|q_\text{L}|^2C^2-|q_\text{R}|^2S^2)\proj{+10}{-01}+(\text{H.c.})_{_{\substack{\text{non-}\\\text{diag.}}}}
\end{align}

Assuming that the observers cannot distinguish between particle and antiparticles yield matrices where only one block of the partial transpose density matrix gives negative eigenvalues
\begin{itemize}
\item Basis: $\{\ket{-10},\ket{+01}\}$
\end{itemize}
\begin{equation}
\frac12\left(\begin{array}{cc}
0& (|q_\text{R}|^2C^2-|q_\text{L}|^2S^2)\\
(|q_\text{R}|^2C^2-|q_\text{L}|^2S^2)&0
\end{array}\right),
\end{equation}

In this case the negativity is given by \[\mathcal{N}^1_{AR}=\frac12\left||q_\text{R}|^2C^2-|q_\text{L}|^2S^2|\right|\]

A similar result is obtained for the system Alice-AntiRob $\rho^1_{A\bar R}$. In this case the only block of the partial transpose that contributes to negativity is
\begin{itemize}
\item Basis: $\{\ket{-10},\ket{+01}\}$
\end{itemize}
\begin{equation}
\frac12\left(\begin{array}{cc}
0& (|q_\text{L}|^2C^2-|q_\text{R}|^2S^2)\\
(|q_\text{L}|^2C^2-|q_\text{R}|^2S^2)&0
\end{array}\right),
\end{equation}
resulting in
\[\mathcal{N}^1_{A\bar R}=\frac12\left||q_\text{L}|^2C^2-|q_\text{R}|^2S^2|\right|.\]

\begin{figure}[h]
\begin{center}
\includegraphics[width=.50\textwidth]{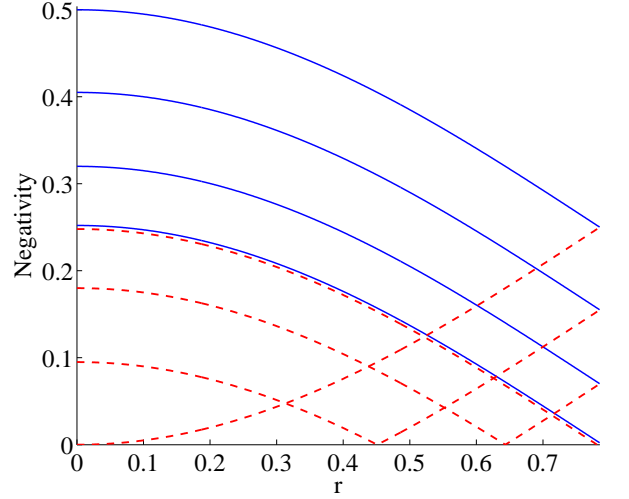}
\end{center}
\caption{Particle-antiparticle maximally entangled state \eqref{3e}: Negativity for the bipartition Alice-Rob (Blue continuous) and Alice-AntiRob (red dashed) as a function of $r_{\Omega}=\arctan{e^{-\pi\Omega/a}}$ for various choices of $|q_\text{\text{R}}|$. The blue continuous (red dashed) curves from top to bottom (from bottom to top) correspond, to $|q_\text{\text{R}}|=1,0.9,0.8,0.7$  respectively. The curves for Alice-AntiRob entanglement have a minimum where the negativity vanishes.}
\label{bundlef}
\end{figure}

Interestingly, when Rob and AntiRob are not able to detect either particle or antiparticle modes the entanglement in the state vanishes. The partial density matrices for $AR$ and $A\bar R$ in this case yield
\begin{align}
\nonumber\rho^1_{AR^+}=&(|q_\text{R}|^2+|q_\text{L}|^2S^2)\proj{+1}{+1}+|q_\text{L}|^2C^2\proj{+0}{+0} \\
\nonumber+&(|q_\text{L}|^2+|q_\text{R}|^2C^2)\proj{-0}{-0}+|q_\text{R}|^2S^2\proj{-1}{-1}\\
+&(\text{H.c.})_{_{\substack{\text{non-}\\\text{diag.}}}}
\end{align}
\begin{align}
\nonumber\rho^1_{A\bar R^-}=&(|q_\text{L}|^2+|q_\text{R}|^2S^2)\proj{+1}{+1}+|q_\text{R}|^2C^2\proj{+0}{+0} \\
\nonumber+&(|q_\text{R}|^2+|q_\text{L}|^2C^2)\proj{-0}{-0}+|q_\text{L}|^2S^2\proj{-1}{-1}\\
+&(\text{H.c.})_{_{\substack{\text{non-}\\\text{diag.}}}}
\end{align}
\begin{align}
\nonumber\rho^1_{AR^+}=&(|q_\text{L}|^2+|q_\text{R}|^2C^2)\proj{+0}{+0}+|q_\text{L}|^2C^2\proj{-0}{-0} \\
\nonumber+&(|q_\text{R}|^2+|q_\text{L}|^2S^2)\proj{-1}{-1}+|q_\text{R}|^2S^2\proj{+1}{+1}\\
+&(\text{H.c.})_{_{\substack{\text{non-}\\\text{diag.}}}}
\end{align}
\begin{align}
\nonumber\rho^1_{A\bar R^+}=&(|q_\text{R}|^2+|q_\text{L}|^2C^2)\proj{+0}{+0}+|q_\text{R}|^2C^2\proj{-0}{-0} \\
\nonumber+&(|q_\text{L}|^2+|q_\text{R}|^2S^2)\proj{-1}{-1}+|q_\text{L}|^2S^2\proj{+1}{+1}\\
+&(\text{H.c.})_{_{\substack{\text{non-}\\\text{diag.}}}}
\end{align}
for which negativity is strictly zero. The entanglement in this state is of a different nature as the entanglement in states $\ket{\Psi_+}$ and $\ket{\Psi_-}$  therefore, a direct comparison of the behaviour of entanglement cannot be done. The total entanglement here is associated to correlations between particles and antiparticles and, therefore, if we trace out either the particles or antiparticles  we effectively remove all the correlations codified in this degree of freedom. 

In the case when the detectors do not distinguish between particles and antiparticles (see Fig. \ref{bundlef}) we found that the entanglement in the Alice-AntiRob bipartition is degraded with acceleration vanishing at a critical point. For higher accelerations entanglement  then begins to grow again. Namely, the entanglement on the bipartition Alice-AntiRob, vanishes for a specific value of the acceleration if $|q_\text{R}|<1$. This value of the acceleration is given by \[\tan^2 r=\frac{1}{|q_\text{R}|^2}-1\]
What is more, the surviving entanglement in the limit $a\rightarrow\infty$ is 
\[\mathcal{N}^1_{AR}(a\rightarrow\infty)=\mathcal{N}_{A\bar R}(a\rightarrow\infty)=\frac14(|q_\text{R}|^2-|q_\text{L}|^2)\]
Therefore,  when $|q_\text{R}|=|q_\text{L}|=1/\sqrt{2}$ no entanglement survives in the limit of infinite acceleration. 

This shows that entanglement has a non-vanishing minimum value in the infinite acceleration limit (regardless the election of Unruh modes) only  when there is transfer of entanglement between particles and antiparticles. Otherwise, it is possible to find an Unruh mode whose entanglement vanishes in the infinite acceleration limit as in the bosonic case. We therefore conclude that the entanglement transfer between particle and antiparticle sectors plays a key role in explaining the behaviour of entanglement in the infinite acceleration limit.

\section{Conclusions}\label{conclusions}

Including antiparticles in the study of fermionic entanglement allowed us to understand key features which explain the difference in behaviour of entanglement  in the fermionic and bosonic case.  Namely, we have shown that there is an entanglement redistribution between the particle and antiparticle sectors when Rob is in uniform acceleration. This entanglement transfer is not possible in the bosonic case and, therefore the differences in the behaviour of entanglement in the bosonic and fermionic case arise. In particular, we have shown that this entanglement tradeoff gives rise to a non-vanishing minimum value of fermionic entanglement in the infinite acceleration limit for any choice of Unruh modes.

We also exhibit a special fermionic state for which entanglement transfer between particle and antiparticle states is not possible. Interestingly, in this case we can find a specific choice of Unruh modes such that entanglement vanishes in the infinite acceleration limit.  Incidentally, this choice ($|q_\text{R}|=|q_{\text{L}}|=1/\sqrt2$) minimises the surviving entanglement of states \eqref{1e} and \eqref{2e}. We showed that it is the tradeoff between the particles and antiparticles sector what protected them from a complete entanglement loss.

Our analysis is based on an extension to antiparticles of the formalism introduced in \cite{Jorma} which relates Unruh and Rindler modes.  This allowed us to analyse a more general family of fermionic maximally entangled states for which the single-mode approximation does not hold.  

This study sheds light in the understanding of  relativistic entanglement: the differences in bosonic and fermionic statistics give rise to differences in entanglement behaviour. This provides a deep insight on the mechanisms which makes fermionic entanglement more resilient to Unruh-Hawking radiation.

\section{Acknowledgements}
We would like to thank J.~Louko and M.~Montero for interesting discussions and helpful comments.  
I. F was supported by EPSRC [CAF Grant EP/G00496X/2]. 
E. M-M was supported by a CSIC JAE-PREDOC2007 Grant, the Spanish MICINN Project FIS2008-05705/FIS  and the QUITEMAD consortium.

\bibliographystyle{apsrev}

\end{document}